\documentclass[mlabstract]{jmlr}

\usepackage{longtable}

\usepackage{booktabs}
\usepackage{nicefrac}

\theorembodyfont{\upshape}
\theoremheaderfont{\scshape}
\theorempostheader{:}
\theoremsep{\newline}

\jmlrvolume{}
\firstpageno{1}
\editors{Sophia Sanborn, Christian Shewmake, Simone Azeglio, Nina Miolane}

\jmlryear{2023}
\jmlrworkshop{Symmetry and Geometry in Neural Representations}

\title[Manifold-augmented Eikonal Equations]{Manifold-augmented Eikonal Equations:\titlebreak Geodesic Distances and Flows on Differentiable Manifolds.}

\author{
  \Name{Daniel Kelshaw}
  \Email{djk21@ic.ac.uk}\\
  \addr{Department of Aeronautics, Imperial College London}
  \AND
  \Name{Luca Magri} \Email{l.magri@imperial.ac.uk}\\
  \addr{Department of Aeronautics, Imperial College London\\ The Alan Turing Institute, British Library}
}

\newcommand\restr[2]{{
  \left.\kern-\nulldelimiterspace 
  #1
  \vphantom{|}
  \right|_{#2}
}}

\DeclareMathOperator*{\argmin}{arg\,min}
\DeclareMathOperator*{\grad}{grad}

\begin{document}

\maketitle

\begin{abstract}
Manifolds discovered by machine learning models provide a compact representation of the underlying data. Geodesics on these manifolds define locally length-minimising curves and provide a notion of distance, which are key for reduced-order modelling, statistical inference, and interpolation. In this work, we propose a model-based parameterisation for distance fields and geodesic flows on manifolds, exploiting solutions of a manifold-augmented Eikonal equation. We demonstrate how the geometry of the manifold impacts the distance field, and exploit the geodesic flow to obtain globally length-minimising curves directly. This work opens opportunities for statistics and reduced-order modelling on differentiable manifolds.
\end{abstract}
\begin{keywords}
Riemannian manifolds, Geodesic, Distance
\end{keywords}

\section{Introduction}
\label{sec:introduction}

The manifold hypothesis states that many high-dimensional systems are expressable on embedded, low-dimensional manifolds~\citep{fefferman2016TestingManifoldHypothesis}. Standard Euclidean measures of distance are insufficient for providing meaningful measures of similarity, and thus distance measures which take into account the geometry of the underlying manifold are necessary. These distance measures are of particular importance for computing statistics on manifolds, or for reduced-order modelling~\citep{guigui2023IntroductionRiemannianGeometry, pennec2006IntrinsicStatisticsRiemanniana}.

\paragraph{Related Work.}

Standard methods compute geodesic distances on the manifold by parameterising and minimising the energy of paths between points~\citep{chen2018MetricsDeepGenerative, kalatzis2020VariationalAutoencodersRiemannian}; however, these methods yield pairwise distances rather than a continuous distance function. Solutions to the Eikonal equation can provide such functions, but prior applications on manifolds are limited. Euclidean methods presented by the likes of~\citet{gropp2020ImplicitGeometricRegularization} do not handle nonlinear manifolds. Works by \citet{crane2013GeodesicsHeatNew, spira2004EfficientSolutionEikonal} develop discrete solutions, which prove intractable in high-dimensions and do not offer differentiable representations. To our knowledge, no existing work solves a manifold-augmented Eikonal equation to obtain a continuous, differentiable represenation of the distance function. Differentiable representations are crucial for obtaining geodesic flows, which characterise globally shortest-paths on the manifold, thus providing an alternative method for computation of geodesics~\citep{carmo2013RiemannianGeometry, lee2018IntroductionRiemannianManifolds}.


\paragraph{Contributions.} 

We first provide a background of material required to specify the Eikonal equation on the manifold in~§\ref{sec:review}. Second, in §\ref{sec:methodology} we describe a method for obtaining solutions to a manifold-augmented Eikonal equation, providing a continuous, differentiable distance function on the manifold. Our proposed approach allows us to obtain length-minimising paths directly. Third, in §\ref{sec:results} we showcase results on a complex manifold with many local extrema. We demonstrate the efficacy of our approach through the successful recovery of globally length-minimising paths. Finally, we discuss limitations and future work in~§\ref{sec:conclusion}.%

\section{Background}
\label{sec:review}

In this section, we provide a brief overview of concepts from differential geometry which form the building blocks of our proposed methodology. Note that, throughout this paper, we employ the Einstein summation convention~\citep{carmo2018DifferentialGeometryCurves, lee2012IntroductionSmoothManifolds}.

\paragraph{Riemannian manifolds and the inner product.}
A Riemannian manifold~\citep{lee2018IntroductionRiemannianManifolds, carmo2013RiemannianGeometry} is a pair $(M, g)$, where $M$ is a smooth manifold \citep{lee2012IntroductionSmoothManifolds}, and $g$ is a choice of Riemannian metric on $M$. This metric constitutes a symmetric bilinear form $g: T_pM \times T_pM \rightarrow \mathbb{R}$, allowing for computation of the inner product on the tangent space,
\begin{equation}
  \langle v, w \rangle_g = g_p(v, w) = g_{ij} v^i w^j.
\end{equation}

\paragraph{Defining the induced metric.}
We consider the case of embedded submanifolds. Suppose $(\tilde{M}, \tilde{g})$ is a Riemannian manifold, and $M \subseteq \tilde{M}$ is an embedded submanifold. Given a smooth immersion $\iota: M \hookrightarrow \tilde{M}$, the metric $g = \iota^\ast \tilde{g}$ is referred to as the metric induced by $\iota$, where $\iota^\ast$ is the pullback. If $(M, g)$ is a Riemannian submanifold of $(\tilde{M}, \tilde{g})$, then for every $p \in M$ and $v, w \in T_p M$, the induced metric is defined as
\begin{equation} \label{eqn:induced_metric}
  g_p \left( v, w \right) 
  = \left( \iota^\ast \tilde{g} \right)_p \left( v, w \right)
  = \tilde{g}_{\iota(p)} \left( d \iota_p(v), d \iota_p(w) \right).
\end{equation}

\paragraph{Defining distance on the manifold.}
\label{sec:review:distance}
While the induced metric $g$ provides the means to compute distances on the tangent space, the notion of distance on the manifold arises from considering the existence of shortest-paths, known as geodesics~\citep{lee2018IntroductionRiemannianManifolds, lee2012IntroductionSmoothManifolds, carmo2018DifferentialGeometryCurves, carmo2013RiemannianGeometry}. These geodesic curves, $\gamma$, satisfy the second-order ordinary differential equation
\begin{equation} \label{eqn:geodesic_equation}
  \nabla_{\dot{\gamma}} \dot{\gamma} = \ddot{\gamma}^k + \Gamma^{k}_{\phantom{k}ij} \dot{\gamma}^i \dot{\gamma}^j = 0,
\end{equation}
where $\nabla$ denotes the covariant derivative, and $\Gamma^k_{\phantom{k}ij}$ are the Christoffel symbols. While geodesics are length-minimising, this is only in a local sense. Given two points $p, q \in M$, the geodesic distance is defined as the infimum of the length of all valid geodesics; precisely
\begin{equation}
  d_g\left( p, q \right) = \inf_\gamma \left\{ \int_0^1 \langle \dot{\gamma}(t), \dot{\gamma}(t) \rangle^{\nicefrac{1}{2}}_{\gamma(t)} dt : \gamma(0) = p, \gamma(1) = q \right\}.
\end{equation}
Conversely, this implies that the shortest curve between two points is itself a geodesic.

\section{Methodology}
\label{sec:methodology}

We propose a new method for computing geodesic distances, theoretically yielding the ability to compute globally length-minimising geodesics directly. We consider a function $\varphi: q \mapsto d_g \left( p, q \right)$, which computes the geodesic distance between points $q, p \in M$. This distance function $\varphi$ satisfies the manifold-augmented Eikonal equation
\begin{equation} \label{eqn:eikonal}
  \lVert \nabla \varphi \rVert
  = \langle \grad \varphi, \grad \varphi \rangle_{g} 
  = \varphi^{,i} \varphi_{,i}
  = 1 
  \qquad \text{s.t.} \quad \restr{\varphi}{p} = 0.
\end{equation}
where $\left( \grad \varphi \right)^i = g^{ij} \varphi_{,j}$ is the exterior derivative; $g^{ij}$ is the inverse of the metric tensor $g_{ij}$; $\varphi_{,i}$ denotes the derivative along the $i$th coordinate vector field; and $\varphi^{,i} = g^{ij} \varphi_{,j}$.  Notably, the exterior derivative of the distance field provides the geodesic flow, a section of the tangent bundle \citep[Thm.~6.31,~6.32]{lee2018IntroductionRiemannianManifolds}. This flow satisfies the geodesic equation \eqref{eqn:geodesic_equation},
\begin{equation} \label{eqn:geodesic_flow}
  \nabla_{\grad \varphi} \grad \varphi = 0,
\end{equation}
where integral paths constitute unit-distance geodesics travelling orthogonal to the level-sets imposed by the distance field. Obtaining the distance function $\varphi_\theta$ yields a method to compute globally length-minimising geodesics directly, alleviating the need to parameterise length-minimising curves between pairs of points.

\paragraph{Neural Network Parameterisation.}
We parameterise the distance function by a neural network $\tilde{\varphi}_\theta : M \rightarrow \mathbb{R}$, providing a continuous, differentiable representation of the solution. The output of the network is augmented to yield the distance function
\begin{equation}
  \varphi_\theta(q; p)
  = \left( \iota^\ast d_E \right)(p, q) \left[ 1 + (\sigma \circ \tilde{\varphi}_\theta) (q; p) \right],
\end{equation}
where $\sigma \in \mathcal{C}^{\geq2} \left( \mathbb{R}, \mathbb{R}_{\geq 0} \right)$ is strictly increasing; and $d_E : (p, q) \mapsto \lVert p - q \rVert_2$ is the Euclidean distance function. Defining the distance function in this manner serves two distinct purposes: \textit{(i)} the constraint $\restr{\varphi}{p} = 0$ is handled implicitly; and \textit{(ii)} the inequality $\varphi(q; p) \geq (\iota^\ast d_E)(p, q)$ is guaranteed~\citep{carmo2013RiemannianGeometry}. We seek a solution ${\varphi_{\theta^\ast} = d_g(q, p)}$ through solving
\begin{equation} \label{eqn:optimisation_problem}
  \theta^\ast = \argmin_\theta{\mathcal{L}(\theta; p)}
    \quad \text{where} \;
  \mathcal{L} = \mathbb{E}_{q} \left( \lVert
    {\varphi_\theta}^{,i}(q) {\varphi_\theta}_{,i}(q) - 1
  \rVert^2 \right),
\end{equation}
where the expectation is taken with respect to some distribution $q \sim \mathcal{D}$ in $M$.

\paragraph{Curvature-based Sampling.}
The solution $\varphi$ to the Eikonal equation is more sensitive in regions of high curvature, a consequence of induced geodesic deviation~\citep{carmo2013RiemannianGeometry}. We mitigate the impact of geodesic deviation through employing a curvature-based sampling strategy, which ensures we can capture the solution appropriately in regions of high curvature. Employing the Metropolis-Hastings~\citep{metropolis1953equation, hastings1970monte} sampling scheme, we draw samples in regions of the manifold that exhibit higher degrees of scalar curvature, which is characterised by the magnitude of the Ricci scalar \citep{lee2018IntroductionRiemannianManifolds}.\footnote{All code is publically available: \url{https://github.com/danielkelshaw/riemax}}

\section{Results}
\label{sec:results}

In order to provide an illustrative example, we demonstrate results for a function $\iota: \mathbb{R}^2 \rightarrow \mathbb{R}^3$, where intrinsic coordinates are interpretable in two-dimensional space. We choose a non-convex manifold with multiple local extrema, exhibiting non-negligible scalar curvature; namely, the peaks function~\citep{mathworksPeaksFunctionMATLABa}. Parameterising the distance function $\varphi_\theta$ with a neural network, we employ the \texttt{adam} optimiser~\citep{kingma2017AdamMethodStochastic}, followed by L-BFGS~\citep{liu1989LimitedMemoryBFGS}, to seek a solution to the optimisation problem \eqref{eqn:optimisation_problem}. Further details on the network architecture and optimisation scheme can be found in Appendix~\ref{apd:network_architecture}.

\begin{figure}[htbp]
\floatconts
  {fig:subfigex}
  {\caption{Results for the distance function, and geodesic flow on the peaks manifold.}}
  {%
    \subfigure[Curvature Sampling]{\label{fig:curvature}%
      \includegraphics[height=0.29\linewidth]{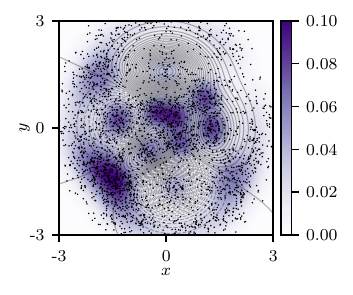}}%
    \hfill
    \subfigure[Geodesic Flow]{\label{fig:geodesics}%
      \includegraphics[height=0.29\linewidth]{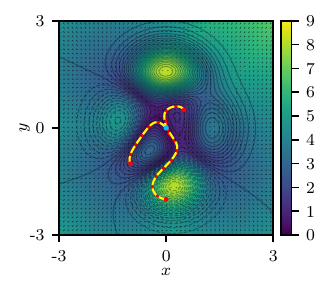}}
    \hfill
    \subfigure[Distance Symmetricity]{\label{fig:distance}%
      \includegraphics[height=0.29\linewidth]{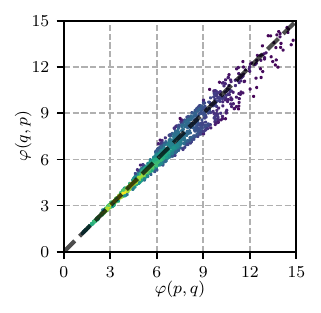}}
  }
\end{figure}

We first use our proposed sampling method, in conjunction with kernel density estimation~\citep{rosenblatt1956RemarksNonparametricEstimates}, to construct a probability density function (PDF) correlated to the magnitude of the scalar curvature; shown in Figure~\ref{fig:curvature}. During optimization, our sample distribution $\mathcal{D}$ is defined by the mean of the curvature PDF and a uniform distribution.

Results in Figure~\ref{fig:geodesics} show the predicted distance field from the centre of the domain $p$, with arrows representing the geodesic flow, $\grad \varphi_\theta$, overlayed. We observe that the geodesic distance adheres to the geometry, showing a tendency to circumvent local extrema where possible. Using the predicted geodesic flow, we obtain length-minimising curves (yellow); these are compared with valid geodesics (red), generated through integration of the geodesic equation \eqref{eqn:geodesic_equation} with $\gamma(0) = q, \dot{\gamma}(0) = \grad \varphi_\theta(q)$. The overlap of these two trajectories demonstrates the validity of the obtained geodesic flow.

Finally, to assess how the choice of $p \in M$ affects the performance, we investigate the symmetricity of the distance function across the domain. We consider a $7 \times 7$ grid of origin points on the manifold, forming the set $\set{X}$. For each $p, q \in \set{X}$, we obtain a distance function and compute pair-wise distances, $\varphi_\theta(q; p), \varphi_\theta(p; q)$, as depicted in Figure~\ref{fig:distance}. We observe that as the distance increases, the variability in predicted distance increases correspondingly; effectively, compounding the error. This result is not unexpected, as distance predictions are naturally reliant on predictions closer to the fixed origin.

\section{Conclusion}
\label{sec:conclusion}
In this work, we address the problem of computing geodesic distance on manifolds. First, we introduce the manifold-augmented Eikonal equation as a means of representing distance functions on the manifold. Second, we propose a model-based parameterisation of this distance function, providing a continuous, differentiable representation with theoretic guarantees. Third, we demonstrate the efficacy of our proposed methodology on a complex manifold, showcasing the ability to recover globally length-minimising paths through leveraging the geodesic flow. Future work will focus on improving robustness as geodesic distance increases, and removing the fixed origin limitation. This work opens opportunities for enhanced distance computation in statistics and reduced-order modelling.

\acks{
D. Kelshaw. and L. Magri. acknowledge support from the UK EPSRC. L. Magri gratefully acknowledges financial support from the ERC Starting Grant PhyCo 949388.
}

\bibliography{references}

\appendix

\section{Network Architecture and Optimisation}\label{apd:network_architecture}
Here, we provide a brief overview of the network architecture and optimisation scheme used throughout this paper.

\paragraph{Network Architecture}
We parameterise the distance function $\varphi_\theta: q \rightarrow d_g(p, q)$ with a neural network $\tilde{\varphi}_\theta$. The architecture of the neural network is characterised by a simple modification of a standard multi-layer perceptron (MLP), as proposed by~\citet{wang2021UnderstandingMitigatingGradient, wang2023ExpertGuideTraining}. For the purposes of this paper, we train the modified MLP with four hidden layers, each with $128$ neurons. The \texttt{tanh} activation is used on all but the output layer. Inputs to the network are standardised according to the domain of interest.

\paragraph{Optimisation Scheme}
We employ a two-stage optimisation scheme: first using the gradient-descent optimiser, \texttt{adam} \citep{kingma2017AdamMethodStochastic}; before finetuning the result with L-BFGS \citep{liu1989LimitedMemoryBFGS}. When using the \texttt{adam} optimiser, we initialise the parameters with standard defaults: $\beta_1 = 0.9, \beta_2 = 0.999, \varepsilon = 10^{-8}, \bar{\varepsilon} = 0$. We define an initial learning rate of $\eta(0) = 10^{-3}$, exponentially decaying by a factor $0.9$ every $2000$ steps. This exponential decay only takes effect after a warm-up period of $5000$ parameter updates, accelerating convergence. We train for a total of $10^{5}$ parameter updates, randomly sampling a total of $2^{13}$ values of $q \in M$ each time. For the second stage of training, we use the L-BFGS optimiser for a further 1000 parameter updates. This approach stabilises training and improves convergence, avoiding local minima. All other training parameters remain the same. Training takes approximately $15$mins on a single NVIDIA RTX 6000 GPU.

\paragraph{Curvature-based Sampling}
We use the Metropolis-Hastings sampling method to generate random samples based on the magnitude of the scalar curvature. In order to ensure the Markov chains are statistically representative, we discard the initial $5\times 10^{4}$ samples. We also draw samples from $N=5$ chains, each with a random initialisation to provide improved coverage of the manifold. In order to generate a probability density function, a kernel density estimate was obtained from the given samples.

\end{document}